\documentclass{raa}

\usepackage[varg]{txfonts}

\usepackage{natbib}
\usepackage{graphicx}
\usepackage{epstopdf}
\usepackage{amssymb}
\usepackage{amsmath,esint}
\usepackage{float}

\bibpunct{(}{)}{;}{a}{}{,}

\begin{document}


\title{A Physical Interpretation of the Titius-Bode Rule and \\ its Connection to the Closed Orbits of Bertrand's Theorem}

   \volnopage{Vol.0 (201x) No.0, 000--000}      
   \setcounter{page}{1}          

\author{
Dimitris M. Christodoulou\inst{1,2}
\and 
Demosthenes Kazanas\inst{3}
}

\institute{
Lowell Center for Space Science and Technology, University of Massachusetts Lowell, Lowell, MA, 01854, USA.\\
\and
Department of Mathematical Sciences, University of Massachusetts Lowell, Lowell, MA, 01854, USA. E-mail: dimitris\_christodoulou@uml.edu \\
\and
NASA Goddard Space Flight Center, Laboratory for High-Energy Astrophysics, Code 663, Greenbelt, MD 20771, USA. Email: demos.kazanas@nasa.gov \\
}

\date{Received~~2017 month day; accepted~~2017~~month day}

\abstract{
We consider the geometric Titius-Bode rule for the semimajor axes of planetary orbits. We derive an equivalent rule for the midpoints of the segments between consecutive orbits along the radial direction and we interpret it physically in terms of the work done in the gravitational field of the Sun by particles whose orbits are perturbed around each planetary orbit. On such energetic grounds, it is not surprising that some exoplanets in multiple-planet extrasolar systems obey the same relation. But it is surprising that this simple interpretation of the Titius-Bode rule also reveals new properties of the bound closed orbits predicted by Bertrand's theorem and known since 1873.
}

\authorrunning{Christodoulou and Kazanas}
\titlerunning{Titius-Bode Rule and Orbits of Bertrand's Theorem}

\maketitle

\section{Introduction}

The numerical algorithm called the Titius--Bode ``law" has been known for 250 years \citep[e.g.,][]{nie72,lec73,dan88,mur99}. It relies on an ad-hoc geometric progression to describe the positions of the planets in the 
solar system and works fairly well out to Uranus but no farther \citep{jak72}.
The same phenomenology has also been applied to the satellites of the 
gaseous giant planets \citep{neu86,mur99}.
Two modern brief reviews of the history along with criticisms of 
this rule have been written by \cite{gra94} and \cite{hay98}.
Currently, the general consensus is that a satisfactory physical basis has not been found for this numerical coincidence despite serious efforts by many researchers over the past three centuries. Furthermore, opinions differ on whether such a physical basis exists at all.

Apparently, many researchers still believe that the Titius--Bode
algorithm does have a physical foundation and continue to work on
this problem. In particular, the last decade of the twentieth century
saw a resurgence of investigations targeting precisely two questions:
the origin of the ``law"  \citep{gra94,dub94,li95,not97,las00} and its statistical robustness against the null hypothesis \citep{hay98,mur99,lyn03}. Furthermore, in this century, some extrasolar systems have been discovered in which the planets appear to obey the Titius-Bode rule and the rule is used as a predictor of additional planets yet to be discovered in these multiple-planet systems \citep{pov08a,pov08b,bov13,hua14,bov15}.

In \S~\ref{work}, we examine the Titius-Bode rule in its original form, that of a geometric progression of the semimajor axes of most of the planetary orbits in the solar system. By inductive reasoning, we associate the geometric rule with the work done in the gravitational field of the Sun by perturbed particles orbiting in the vicinity of planetary orbits, but we find that the spacing of the semimajor axes is not the right qualifier of the physical profile dictated by the Sun's gravitational potential. Then we derive another rule for a group of hypothetical orbits that are equally spaced between the actual semimajor axes and we interpret this rule physically in terms of the gravitational potential differences of particles perturbed around the actual orbits of the planets. Our results support the discovery of \cite{las00} \citep[for related recent works see][]{jia15,las17} that such an arrangement of orbits implies that the protoplanets do not interfere with one another during their formation stage, thus a planet is expected to be formed at every available orbit of the geometric progression. Furthermore, our results reveal new geometric properties (see the Appendix) of the bound closed orbits predicted in spherical potentials by the celebrated theorem of \cite{ber73}. In \S~\ref{sum}, we summarize and discuss these results.

\section{Titius-Bode Rule Rewritten and Interpreted Physically}\label{work}

\begin{figure}
\includegraphics[trim=0 2.5cm 0 0, clip, angle=0,width=18cm]{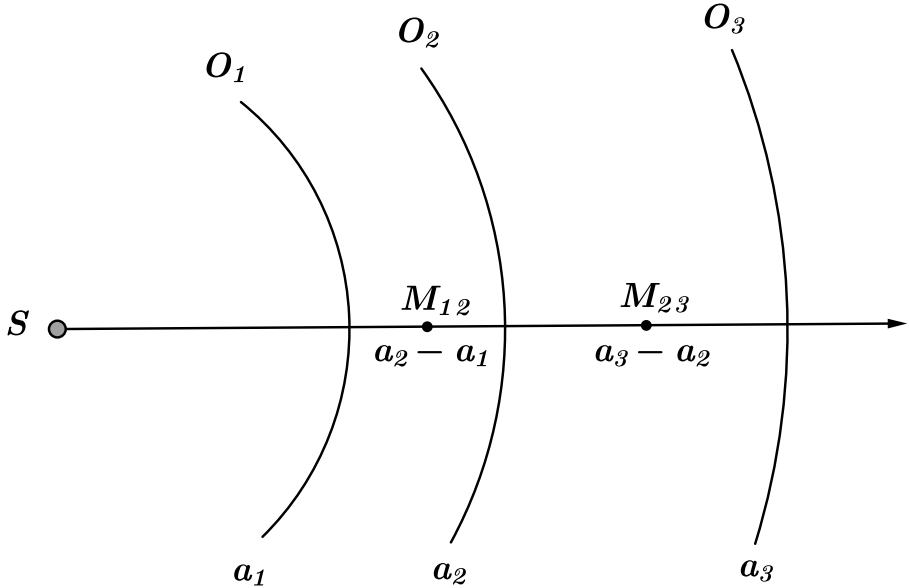}
\caption{Schematic diagram of three consecutive planetary orbits with semimajor axes $a_1$, $a_2$, and $a_3$ in geometric progression ($a_2=\sqrt{a_1 a_3}$). The midpoints $M_{12}$ and $M_{23}$ of the segments between the orbits are also marked along the ray from the Sun $S$.}
\label{fig1}
\end{figure}

In its original form, the Titius-bode rule dictates that the semimajor axes of most planetary orbits are in geometric progression. (In some forms, an additional term of 0.4 is added ad hoc in order to reproduce the innermost three planets that appear to be in arithmetic progression.)
The geometric progression is described formally by two equivalent relations: Consider three consecutive orbits with semimajor axes $a_1$, $a_2$, and $a_3$ (Fig.~\ref{fig1}); then the intermediate axis must be the geometric mean of its neighboring axes, viz.
\begin{equation}
a_2 = \sqrt{a_1 a_3}\, ,
\label{tb1}
\end{equation}
or equivalently
\begin{equation}
\frac{1}{a_2-a_1} - \frac{1}{a_3-a_2} = \frac{1}{a_2}\, . 
\label{tb2}
\end{equation}
The form of eq.~(\ref{tb2}) contains reciprocal distances and this is a sufficient hint that the relation could be associated with the central gravitational potential due to the Sun. But, as illustrated in Figure~\ref{fig1}, such a simple association is not entirely straighforward because the distances $(a_2-a_1)$ and $(a_3-a_2)$ are not central, i.e., they are not measured from the Sun. 

In order to recast the rule in terms of central reciprocal distances, we define hypothetical orbits that are equidistant between the semimajor axes. In Figure~\ref{fig1}, such orbits would cross the ray from $S$ at the midpoints $M_{12}$ and $M_{23}$. Their radial coordinates are
\begin{equation}
m_{12} = \frac{1}{2}(a_1+a_2) \ \ \ \ {\rm and} \ \ \ \ m_{23} = \frac{1}{2}(a_2+a_3) \, ,
\label{tb3}
\end{equation}
respectively. The sequence $m_{12}, m_{23}, ...$ of intermediate radii forms a geometric progression with the same ratio as that of the $a_1, a_2, a_3, ...$ sequence. Eliminating $a_1$ and $a_3$ between eqs.~(\ref{tb1}) and~(\ref{tb3}), the Titius-Bode rule is transformed to the equivalent form
\begin{equation}
\frac{1}{m_{12}} + \frac{1}{m_{23}} = \frac{2}{a_2}\, ,
\label{tb4}
\end{equation}
which implies that $a_2$ is the harmonic mean of $m_{12}$ and $m_{23}$. As we describe in the Appendix, this is an important geometric property that is valid only in a central $-1/r$ gravitational potential and its physical meaning can be easily deduced:
eq.~(\ref{tb4}) can be rewritten in a form that can be interpreted in terms of central potential differences, viz.
\begin{equation}
{\cal G M}\left(\frac{1}{m_{12}} - \frac{1}{a_2}\right) \ = \ {\cal G M}\left(\frac{1}{a_2} - \frac{1}{m_{23}}\right) \, ,
\label{tb5}
\end{equation}
where ${\cal G}$ is the gravitational constant and ${\cal M}$ is the mass of the central object that creates the gravitational field.

Consider now particles oscillating about the intermediate orbit $O_2$ (this includes also the protoplanetary core early in its formation and before it settles down to $O_2$).
It is evident that the work done by a particle at $m_{12}$ to reach $a_2$ is the same as the work done by the field to a particle at $m_{23}$ that reaches $a_2$.
In other words, the gravitational field allows orbit $O_2$ in Figure~\ref{fig1} to utilize the entire area between the hypothetical orbits through $M_{12}$ and $M_{23}$ and to accummulate matter while sharing half-way with orbits $O_1$ and $O_3$ the areas between them. This arrangement of orbits in a geometric progression ensures that adjacent orbits do not interfere with one another, a result that was first found by \cite{las00} who started with intersecting planitesimal orbits and derived the Titius-Bode rule for a surface density profile of the solar nebula when the interactions ceased. Our derivation above starts with the Titius-Bode rule and it is effectively the converse of Laskar's derivation.

This ``harmonic-mean'' sharing by protoplanets of the in-between areas has also been used empirically in the seminal work of \cite{wei77} who distributed planetary material in annuli around the current orbits of planets in order to derive a surface density profile for the solar nebula. Our calculation justifies this empirical notion on energetic grounds: eq.~(\ref{tb5}) describes the energy balance of a harmonic oscillator in spherical (radial) coordinates with different amplitudes on either side of orbit $O_2(a_2)$ and it is in contrast to the simple harmonic oscillator in which the deviations $(a_2-a_1)/2$ and $(a_3-a_2)/2$ from the equilibrium position $a_2$ are equal because of the linear nature of the restoring force \citep{hoo1678}.

\section{Summary and Discussion}\label{sum}

\subsection{Summary}

We have described a physical interpretation of the Titius-Bode rule by considering, not the present positions of the planets in the solar system, but the ``regions of occupancy'' utilized by neighboring protoplanets during their efforts to collect and accummulate material as they orbit in the solar nebula: according to eq.~(\ref{tb5}), the work done by a particle to move out from an interior orbit through $M_{12}$ (Fig.~\ref{fig1}) to the next outer planetary orbit $O_2$ is the same as the work done to a particle that falls into the gravitational field from $M_{23}$ to $O_2$. 

The importance of protoplanets sharing half-way their in-between regions is twofold. First, the protoplanets do not cross into the orbits of their neighbors as they oscillate about their equilibrium orbits and continue to accummulate material \citep{las00,jia15,las17}. This behavior ensures that some object or objects will be found in every single radial location $a_1, a_2, a_3, ...$, even in the predicted location between Mars and Jupiter (where the asteroid belt resides). Second, after the remaining disk gas disperses or gets accreted by the Sun and the planets emerge in their final settled orbits, the long-term dynamical stability of the solar system is strengthened because these orbits are as far away from one another as possible, and neighboring planets may interact only weakly by tidal forces that exert only minor perturbations to the positions of their neighbors \citep{hay98}. Such weak interactions are contingent upon the absence of resonant orbits which is an observed fact for the planets in our solar system.

\subsection{Solar Nebula}

In \cite{chr07}, we derived exact solutions of the Lane-Emden equation with rotation for the solar nebula \citep{lan1870,emd1907} assuming it is an isothermal gas. The isothermal solutions of the Lane-Emden equations are very much relevant to the problem at hand: they show that protoplanetary cores are trapped inside local gravitational potential wells in which they can collect matter and grow in time. The distances of these localized potential wells from the protosun are in geometric progression as a result of the differential rotation of the solar nebula (that tapers off at the inner region and at the farthest outer regions of the nebula, where the planetary orbits appear to follow arithmetic progressions). 

The present result comes to strengthen the argument that planets grow locally inside deep gravitational potential wells that extend half-way between adjacent planetary orbits: on energetic grounds, solid protoplanetary cores share the disk space in the solar nebula between adjacent orbits and they collect material by various processes that make matter settle down to the potential minima, whereas the gas can flow inward and continue its accretion on to the central protosun. Furthermore, this model argues against excessively large migrations of protoplanets in the solar nebula \citep[][and references therein]{gom04,gom05,lev07}. Protoplanetary cores can move radially only within the bounds of their local gravitational potential wells (radii $m_{12}$ and $m_{23}$ in eq.~(\ref{tb3}) for orbit $O_2$ in Fig.~\ref{fig1}).

\subsection{Extrasolar Multiplanet Systems}

It is not surprising that at least some extrasolar systems exhibit similar characteristic distributions of exoplanetary orbits. Their protoplanetary disks may have had similar energetic and stability properties as our solar nebula, a similarity that apparently is neither universal nor wide-spread \citep{hua14,bov15}. As for the location of the habitable zone and its planets in extrasolar systems \citep{kan16}, we believe that the outcome depends crucially on the differential rotation and surface density profiles of each particular protoplanetary disk \citep{las00,chr07,jia15} irrespective of whether the Titius-Bode rule is applicable or not.

\subsection{Connection to the Closed Orbits of Bertrand's Theorem}\label{sec_bertrand}

Eq.~(\ref{tb5}) shows that perturbed particle orbits around a circular equilibrium orbit such as $O_2(a_2)$ in Figure~\ref{fig1} have different amplitudes, say $A_1$ and $A_2>A_1$, on either side of the equilibrium radius $a_2$. This is required so that the potential differences between $a_2$ and the maximum radial displacements be equal in magnitude, an assertion of the Work-Energy Theorem 
between the equilibrium radius $a_2$ and the radii of the turning points of the oscillation where the radial velocity goes to zero. The result is a restriction placed on the two amplitudes that must be related by
\begin{equation}
\frac{1}{a_2-A_1} + \frac{1}{a_2+A_2} = \frac{2}{a_2} \, ,
\label{constraint1}
\end{equation}
that is, radius $a_2$ is the {\it harmonic mean} of the radii of the turning points. This property is valid only for bound closed orbits in a $-1/r$ gravitational potential and it is derived in the Appendix, where we also analyze closed orbits in an $r^2$ gravitational potential \citep{ber73}. It turns out that the latter orbits exhibit another precise symmetry altogether: radius $a_2$ is the {\it geometric mean} of the radii of the turning points.

\begin{acknowledgements}
We thank the reviewers of this article for their comments that led to a clearer presentation of our ideas. DMC is obliged to Joel Tohline for advice and guidance over many years.
\end{acknowledgements}

\setcounter{equation}{0}
\renewcommand{\theequation}{A\arabic{equation}}

\section*{Appendix A: The Geometry of Bound Closed Orbits in Spherical Potentials}

\subsection*{A1. Newton-Kepler $-1/r$ Potential}

Consider an equilibrium orbit $r=a$ in a $-1/r$ potential and assume that the maximum radial deviation is $\pm A$ on either side of $r=a$. At the turning points $r=a\pm A$, the radial velocity is zero ($\dot{r}=0$) and the total energy per unit mass can then be written as \citep{gol50}
\begin{equation}
{\cal E} = \frac{{\cal L}^2}{2r^2} - \frac{{\cal GM}}{r} \, ,
\label{en0}
\end{equation}
where the specific angular momentum satisfies ${\cal L}^2 = {\cal GM}a$, thus eq.~(\ref{en0}) can be written in the form
\begin{equation}
\frac{\cal E}{\cal GM} = \frac{a}{2r^2} - \frac{1}{r} = {\rm const.}
\label{en1}
\end{equation}
Applied to the turning points $r=a\pm A$, this equation yields
\begin{equation}
\frac{a}{2(a-A)^2} - \frac{1}{a-A} = \frac{a}{2(a+A)^2} - \frac{1}{a+A} \, ,
\label{en2}
\end{equation}
a strict requirement for energy conservation. This requirement is satisfied only for $A=0$ which implies that the amplitude of the oscillation cannot be the same on either side of $r=a$. 

We consider next two different amplitudes $A_1>0$ and $A_2>A_1$ on either side of the equilibrium orbit $r=a$. After some elementary algebra, energy conservation (eq.~(\ref{en1})) at the turning points $r=a-A_1$ and $r=a+A_2$ yields 
\begin{equation}
\frac{1}{A_1} - \frac{1}{A_2} = \frac{2}{a} \, ,
\label{en3}
\end{equation}
or equivalently
\begin{equation}
\frac{1}{a-A_1} + \frac{1}{a+A_2} = \frac{2}{a} \, .
\label{en3b}
\end{equation}
This last equation shows that, in a $-1/r$ potential, the equilibrium radius $a$ is the harmonic mean of the radii of the turning points $a-A_1$ and $a+A_2$ (as was also found in eq.~(\ref{tb4}) for orbit $O_2$ and points $M_{12}, M_{23}$ in Fig.~\ref{fig1}).

\subsection*{A2. Isotropic Hooke $r^2$ Potential}

The isotropic harmonic-oscillator potential, written as $\Omega^2 r^2/2$ ($\Omega=$const.), cannot support arbitrarily large oscillations of equal amplitude on either side of the equilibrium orbit $r=a$ either. The same analysis leads to an energy equation analogous to eq.~(\ref{en1}), but here ${\cal L}^2=\Omega^2 a^4$, thus
\begin{equation}
\frac{\cal E}{\Omega^2/2} = \frac{a^4}{r^2} + r^2 = {\rm const.}
\label{en4}
\end{equation}
When energy conservation is applied between the turning points $r=a\pm A$, we obtain three solutions, $A=0$ and two extraneous solutions $A=\pm a\sqrt{2}$. The solution $A=a\sqrt{2}$ is of course rejected because $A>a$. 

We consider next two different amplitudes $A_1>0$ and $A_2>A_1$ on either side of the equilibrium orbit $r=a$. After some elementary algebra, energy conservation (eq.~(\ref{en4})) at the turning points $r=a-A_1$ and $r=a+A_2$ yields
\begin{equation}
\frac{1}{A_1} - \frac{1}{A_2} = \frac{1}{a} \, ,
\label{en5}
\end{equation}
or equivalently
\begin{equation}
(a-A_1)(a+A_2) = a^2 \, .
\label{en5b}
\end{equation}
This last equation shows that, in a harmonic $r^2$ potential, the equilibrium radius $a$ is the geometric mean of the radii of the turning points $a-A_1$ and $a+A_2$.



\begin{thebibliography}{}

\bibitem[Bertrand(1873)]{ber73}
Bertrand, J. 1873, C. R. Acad. Sci. Paris, 77, 849

\bibitem[Bovaird \& Lineweaver(2013)]{bov13}
Bovaird, T., \& Lineweaver, C. H. 2013, \mnras, 448, 3608

\bibitem[Bovaird et al.(2015)]{bov15}
Bovaird, T., Lineweaver, C. H., \& Jacobsen, S. K. 2015, \mnras, 448, 3608

\bibitem[Christodoulou \& Kazanas(2007)]{chr07}
Christodoulou, D. M., \& Kazanas, D. 2007, arXiv:0706.3205

\bibitem[Christodoulou \& Kazanas(2017)]{chr17b}
Christodoulou, D. M., \& Kazanas, D. 2017, RAA, submitted



\bibitem[Danby(1988)]{dan88}
Danby, J. M. A. 1988, Fundamentals of Celestial Mechanics, 2nd edition (Richmond: Willmann-Bell)


\bibitem[Dubrulle \& Graner(1994)]{dub94}
Dubrulle, B., \& Graner, F. 1994, A\&A, 282, 269

\bibitem[Emden(1907)]{emd1907}
Emden, R. 1907, Gaskugeln (Leipzig: B. G. Teubner)

\bibitem[Goldstein(1950)]{gol50}
Goldstein, H. 1950, Classical Mechanics (Reading, MA: Addison-Wesley)

\bibitem[Gomes et al.(2004)]{gom04}
Gomes, R. S., Morbidelli, A., \& Levison, H. F. 2004, Icarus, 170, 492

\bibitem[Gomes et al.(2005)]{gom05}
Gomes, R. S., Gallardo, T., Fern\'andez, J. A., \& Brunini, A. 2005, 
Celest. Mech. Dyn. Astron., 91, 109

\bibitem[Graner \& Dubrulle(1994)]{gra94}
Graner, F., \& Dubrulle, B. 1994, A\&A, 282, 262

\bibitem[Hayes \& Tremaine(1998)]{hay98}
Hayes, W., \& Tremaine, S. 1998, Icarus, 135, 549

\bibitem[Hooke(1678)]{hoo1678}
Hooke, R. 1678, De Potentia Restitutiva, or of Spring. Explaining the Power of Springing Bodies (London: J. Martyn)

\bibitem[Huang \& Bakos(2014)]{hua14}
Huang, C. X., \& Bakos, G. \'A. 2014, \mnras, 442, 674

\bibitem[Jaki(1972)]{jak72}
Jaki, S. 1972, Am. J. Phys., 40, 93

\bibitem[Jiang et al.(2015)]{jia15}
Jiang, I.-G., Yeh, L.-C., \& Hung, W.-L. 2015, \mnras, 449, L65

\bibitem[Kane et al.(2016)]{kan16}
Kane, S. R., Hill, M. L., Kasting, J. F., et al. 2016, \apj, 830, 1 


\bibitem[Lane(1870)]{lan1870}
Lane, L. J. H. 1870, Amer. J. Sci. Arts, Second Series, 50, 57

\bibitem[Laskar(2000)]{las00}
Laskar, J. 2000, Phys. Rev. Lett., 84, 3240

\bibitem[Laskar \& Petit(2017)]{las17}
Laskar, J., \& Petit, A. 2017, A\&A, in press (arXiv:1703.07125) 

\bibitem[Lecar(1973)]{lec73}
Lecar, M. 1973, Nature, 242, 318

\bibitem[Levison et al.(2007)]{lev07}
Levison, H. F, Morbidelli, A., Gomes, R. S., \& Backman, D. 2007,
in Protostars and Planets~V, ed. B. Reipurth, D. Jewitt, \& K. Keil (Tucson: Univ. of Arizona Press), 669

\bibitem[Li et al.(1995)]{li95}
Li, X. Q., Zhang, H., \& Li, Q. B. 1995, A\&A, 304, 617

\bibitem[Lynch(2003)]{lyn03}
Lynch, P. 2003, \mnras, 341, 1174

\bibitem[Murray \& Dermott(1999)]{mur99}
Murray, C. D., \& Dermott, S. F. 1999, Solar System Dynamics (Cambridge, UK: Cambridge Univ. Press)

\bibitem[Neuh{{\"a}}user \& Feitzinger(1986)]{neu86}
Neuh{{\"a}}user, R., \& Feitzinger, J. 1986, A\&A, 170, 174

\bibitem[Nieto(1972)]{nie72}
Nieto, M. M. 1972, The Titius--Bode Law of Planetary Distances: Its History and Theory (Oxford, UK: Pergamon Press)

\bibitem[Nottale et al.(1997)]{not97}
Nottale, L., Schumacher, G., \& Gay, J. 1997, A\&A, 322, 1018

\bibitem[Poveda \& Lara(2008a)]{pov08a}
Poveda, A., \& Lara, P. 2008a, Rev. Mex. Ast. Astrof., 34, 49

\bibitem[Poveda \& Lara(2008b)]{pov08b}
Poveda, A., \& Lara, P. 2008b, Rev. Mex. Ast. Astrof., 44, 243

\bibitem[Weidenschilling(1977)]{wei77}
Weidenschilling, S. J. 1977, A\&SS, 51, 153

\end{thebibliography}
\end{document}